\documentclass[journal,10pt,twocolumn]{IEEEtran}
\usepackage{epsfig}
\usepackage{graphicx}
\usepackage{subfigure}

\newtheorem{proposition}{Proposition}

\usepackage{amssymb}
\usepackage{makeidx}
\usepackage{amsmath}
\usepackage{graphicx}
\usepackage{epsf}
\usepackage{epsfig}
\usepackage{ccaption}
\usepackage{array}
\usepackage{tabularx}
\usepackage{multirow}
\usepackage{epsfig}
\usepackage{cite}

\begin{document}
\topmargin=-0.6in \oddsidemargin -0.5in \textwidth=7.3in
\textheight=9.5in

\title{\LARGE{Cooperative Transmission Protocols with High Spectral Efficiency and High Diversity Order Using Multiuser Detection and Network Coding}}
\author{Zhu Han$^+$, Xin Zhang, and H. Vincent Poor\\
$^+$Department of Electrical and Computer Engineering, Boise State
University, Idaho, USA\\
School of Engineering and Applied Science, Princeton University, New
Jersey, USA. \vspace{-10mm} \thanks{This research was supported by
the National Science Foundation under Grants ANI-03-38807,
CCR-02-05214 and CNS-06-25637.}}

\maketitle

\begin{abstract}
Cooperative transmission is an emerging communication technique
that takes advantages of the broadcast nature of wireless
channels. However, due to low spectral efficiency and the
requirement of orthogonal channels, its potential for use in
future wireless networks is limited. In this paper, by making use
of multiuser detection (MUD) and network coding, cooperative
transmission protocols with high spectral efficiency, diversity
order, and coding gain are developed. Compared with the
traditional cooperative transmission protocols with single-user
detection, in which the diversity gain is only for one source
user, the proposed MUD cooperative transmission protocols have the
merits that the improvement of one user's link can also benefit
the other users. In addition, using MUD at the relay provides an
environment in which network coding can be employed. The coding
gain and high diversity order can be obtained by fully utilizing
the link between the relay and the destination. From the analysis
and simulation results, it is seen that the proposed protocols
achieve  higher diversity gain, better asymptotic efficiency, and
lower bit error rate, compared to traditional MUD and to existing
cooperative transmission protocols.

\end{abstract}\vspace{-7mm}


\section{Introduction}\label{sec:intro}

Cooperative transmission \cite{bib:Aazhang1, bib:Laneman2} is a
new communication technique that takes advantage of the broadcast
nature of wireless channels. Recent work has explored cooperative
transmission in a variety of
 scenarios, including cellular networks \cite{bib:ICC06}, ad
hoc/sensor networks \cite{bib:WCNC06_sensor}, and ultra wide band
\cite{bib:WCNC06_UWB}. One drawback of existing cooperative transmission
schemes is
a consequent reduction of spectral efficiency. Moreover, most existing
techniques require orthogonal channels, which are   not
available for many wireless networks such as 3G cellular
networks.

Multiuser Detection (MUD) \cite{verdu} deals with the demodulation of
mutually interfering digital streams, exploiting the
cross-correlations among users to produce better detection performance
in the presence of multiple-access interference. MUD implementations
for cellular applications have been developed by Datang Telecommunication for
TD-SCDMA and by Qualcomm for EVD0. Recently, \cite{HuaiyuDai} has considered the
joint optimization of Multiple Input Multiple
Output (MIMO) systems with MUD. Since
a relay in a cooperative communication scheme
can be viewed as a virtual antenna for a source node, this work motivates
the study of MUD performance in cooperative transmission.
However, unlike MIMO MUD where all information from different antennas can
be obtained without limitation, in cooperative communication schemes
the link between the relay (i.e., the
virtual antenna) and the destination is limited.

To overcome this limitation, network coding
\cite{network_coding, network_coding_coop} provides a potential
solution. The core notion of network coding is to allow mixing of
data at the intermediate network nodes, to improve the overall
reliability of transmission across the network. A receiver receives these
mixed data packets from various nodes and deduces from them the messages that were
originally intended for that data sink. In cooperative
transmission, the relay can be viewed as an intermediate network
node. If MUD is employed at the relay, the relay can employ
network coding by mixing some users' data and transmitting them
through the limited link to the destination node. At the destination,
the performance of all users destined to
that node can be
greatly improved by decoding this coded data.  This issue is examined
in the present paper.

In particular, we propose two cooperative transmission protocols
that make use of MUD and network coding. In the first protocol, realizing
that improvement in one user's decoding can help the decoding
of the other users, we
decide which relays to use and whose information the selected
relays will retransmit such that the overall system performance can be
optimized. In the second protocol, we assume the relays
are equipped with MUD. Then the selected users' information is
coded by network coding and is relayed to the base station. At the
base station, the coding gain is not only realized for the selected users
but also for the other users because of the MUD. From both analytical and
simulation results, it is seen that the proposed protocols achieve
higher diversity and coding gain, better asymptotic efficiency,
and lower bit error rate (BER) than existing schemes.

This paper is organized as follows. In Section \ref{sec:model},
system models are given for multiuser cooperative transmission
with MUD in a base station and a set of relays. In Section
\ref{sec:Protocol}, the two above-noted protocols are constructed.
In Section \ref{sec:analysis}, the properties of the proposed
protocols are studied. Simulation results are shown in Section
\ref{sec:simulation}, and conclusions are drawn in Section
\ref{sec:conclusion}.\vspace{-3mm}

\section{System Model}\label{sec:model}

\begin{figure}[htbp]
\begin{center}
    \epsfig{file=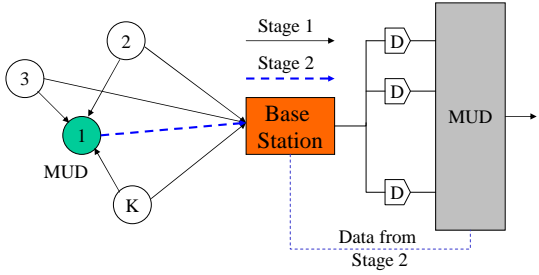,width=71truemm}
\end{center}
\caption{System Model}\label{system_model}\vspace{-3mm}
\end{figure}

Consider a wireless network with  $K$ synchronized uplink users
(i.e., terminals). Among these terminals, $N$  can  serve as
relays. This system model is illustrated with $N=1$ in Figure
\ref{system_model}. At a first transmission stage, all users
except the relays send information, and the relays listen (and
perform MUD if they have the ability). At a second stage, the other
users send their next information signals, while the relays send a
certain user's information or key information from the results of
MUD applied at the first stage. In the base station, all of the
other users' information from the first stage is delayed for one
time slot and jointly decoded with the key information sent by the
relays at the second stage. Since the users cannot transmit and
receive at the same time or at the same frequency, to relay once
costs at least two time slots for listening and relaying. So the
spectral efficiency is $\frac {K-2N} K$, and thus when the number
of users is much larger than the number of relays the spectral
efficiency approaches 1.

We consider an uplink synchronous CDMA system with Gaussian ambient noise.
Define $\mathfrak{R}$ as the group of
relay terminals, and $\mathfrak{L}$ as the group of terminals listening and
preparing for a relay in the next time slot. The received signal at
the base station can be expressed as
\begin{equation}
y(t)=\sum_{k\in K\backslash \mathfrak{R}  \backslash \mathfrak{L}
} A_k b_k s_k(t)+ \sum_{k\in  \mathfrak{R}  } A_kz_ks_k(t)+ \sigma
n(t),
\end{equation}
and at user $i\in\mathfrak{L}$, who is listening and preparing for a
relay in the next time slot, as
\begin{equation}
y^i(t)=\sum_{k\in K\backslash \mathfrak{R}  \backslash
\mathfrak{L} } A_k^i b_k s_k(t)+ \sum_{k\in  \mathfrak{R}  } A_k^i
z_k s_k(t)+\sigma ^i n^i(t),
\end{equation}
where $A_k$ is the received amplitude of the $k^{th}$ user's
signal at the base station, $A_k^i$ is the received amplitude of
the $k^{th}$ user's signal at relay $i$, $b_k\in \{ -1,+1\}$ is
the data symbol transmitted by the $k^{th}$ user, $z_k$ is the relayed
bit, $s_k$ is the unit-energy signature waveform of the $k^{th}$
user, $n(t)$ and $n^i(t)$ are normalized white Gaussian noise processes,
and $\sigma^2$ and $(\sigma ^i)^2$ are  background noise power
densities. For simplicity, we assume
$\sigma=\sigma^i$, although the more general case is straightforward.

The received signal vectors at the base station and at the relay
after  processing by a  matched filter bank can be written as:
\begin{equation}
\textbf y=\textbf{R}\textbf{A}\textbf{b}+\textbf n,
\end{equation}
and
\begin{equation}
\textbf y ^i=\textbf{R}\textbf{A}^i\textbf{b}+\textbf n ^i,
\end{equation}
where $\textbf A=diag\{A_1,\dots , A_{K}\}$, $\textbf
A^i=diag\{A^i_1,\dots , A^i_{K}\}$, $E[\textbf n \textbf
n^T]=E[\textbf n^i \textbf n^{iT}]=\sigma^2\textbf R$, $\textbf R$
is the cross-correlation matrix, whose ${i-j}^{\rm th}$ element
can be written
as
\begin{equation}
    int _{0}^T s_i(t)s_j(t)dt,
\end{equation}
where $T$ is the inverse of the data rate, and $\textbf b=[b_1,
\dots, z_i, \dots, 0, b_{K}]^T$  consists of symbols of direct-transmission,
relay, and listening users.

In this paper, we will investigate the BER performance of MUD
under cooperative transmission. Specifically, we will consider the
optimal MUD and successive cancellation detector which is one type
of decision driven MUD. As pointed out in \cite{verdu}, there is
no explicit expression for the error probability of the optimal
multiuser detector, and bounds must be used. A tight upper
bound is provided by the following proposition from \cite{verdu}.
\begin{proposition}
The BER of the $i^{th}$ user for optimal MUD is bounded according to
\begin{equation}
P_r^{i,opt}\le \sum_{\mathbf{\epsilon} \in F_i}
2^{-\omega(\mathbf{\epsilon})} Q\left(
\frac{\|S(\mathbf{\epsilon})\|}{\sigma} \right)
\end{equation}
where $\mathbf{\epsilon}$ is a possible error vector for user $k$,
and $\|S(\mathbf{\epsilon})\|^2
=\mathbf{\epsilon}^T\mathbf{H}\mathbf{\epsilon}=
\mathbf{\epsilon}^T\mathbf{ARA}\mathbf{\epsilon}$.
$\omega(\mathbf{\epsilon})$ is the number of nonzero elements in
$\mathbf{\epsilon}$, and $F_i$ is the subset of indecomposable
vectors. Due to limited space, for details refer to \cite[Chapter
4]{verdu}.
\end{proposition}

For the successive cancellation detector, a recursive approximation
 is given by the following
proposition\cite{verdu}.
\begin{proposition}
The BER of the $i^{th}$ user for successive cancellation MUD is
given approximately by
\begin{equation}\label{BER_SC}
P_r^{i,sc}\approx Q \left( \frac {A_i}{\sqrt{\sigma^2+\frac 1 M
\sum_{j=1}^{i-1}A_j^2+\frac 4 M \sum_{j=i+1}^K
A_j^2P_j^{sc}}}\right) ,
\end{equation}
where $M$ is the spreading gain.
\end{proposition}

Notice that the BER is a function of the received amplitudes of
the users, which in turn are functions of the user locations and
the network topology. As will be shown later, we have degrees of
freedom to select which users will serve as relays and which users'
information to relay, so as to achieve optimal performance in
terms of BER at the base station.

%
%
%
%

\vspace{-3mm}

\section{Two Cooperative Transmission Protocols\label{sec:Protocol}}


The first protocol seeks to exploit the fact that MUD can improve
all signals to be decoded better because of the mitigation of
interference from the strongest signals. Suppose terminal $i$ is
selected as the relay and it forwards user $m$'s information. At
the base station, after a matched filter bank, maximal ratio
combining (MRC) is used to combine the signals from these two
terminals. The resulting SINR is given by:
\begin{equation}\label{MRC}
\Gamma_m=\Gamma_m^0+\Gamma_i^0,
\end{equation}
where $\Gamma_m^0$ and $\Gamma_i^0$ are the SINRs  to the base station
from user $m$
and user $i$, respectively. Since the optimal
and decision driven MUD algorithms are nonlinear, a closed form
expression for  (\ref{MRC}) is not available. Moreover, the noise
terms corrupting different users after MUD are correlated so that
(\ref{MRC}) can serve only to provide a performance upper bound. In
this paper, we assume that some method such as a threshold test
\cite{bib:Laneman2} is employed so that the potential relays and
the base station can know whether the detected signals are
correct. Also, rather than using
 MRC before the decoding, the final decision is
based on the decoded signals in both stages. An error occurs only
if the decisions in both stages are wrong. So the probability of
error can be written as
\begin{equation}\label{protocol_one_BER}
P_r^m=P_r^{m0}(1-(1-P_r^{mi})(1-P_r^{i0})).
\end{equation}
Here user $i$ is selected as the relay. The error probabilities
for user $m$ to the base station, user $i$ to the base station,
and for user $m$ to user $i$ are denoted as $P_r^{m0}$,
\begin{figure}[htbp]
\begin{center}
    \epsfig{file=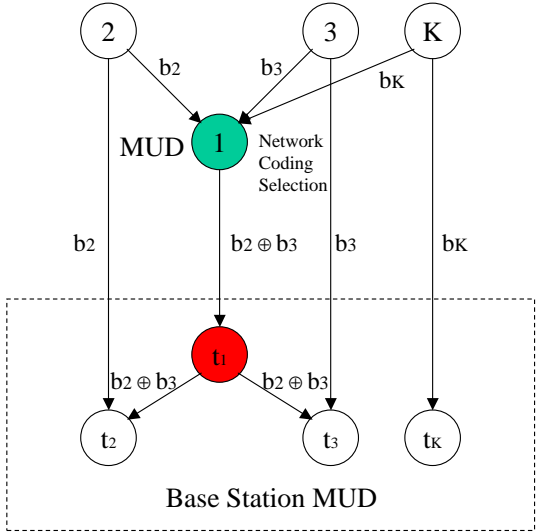,width=71truemm}
\end{center}
\caption{Joint MUD and Network Coding\vspace{-5mm}
Example}\label{network_coding_model}
\end{figure}$P_r^{i0}$, and $P_r^{mi}$, respectively. Notice that there is no
need for MUD at the relays for the first protocol.
%

The second protocol seeks to exploit the fact that MUD in the base
station and the relay provides a possible data-flow structure for
jointly optimizing MUD and network coding. In Figure
\ref{network_coding_model}, we illustrate an example where there
are $K$ users and user $1$ is assigned as the relay. At a first
stage, users $2$ through $K$ send their own information, while the
base station and user $1$ listen. At a second stage, user $1$
sends the coded information (here $b_2 \bigoplus b_3$). Then the
base station can improve the decoding of user $2$ and user $3$.

In general, we can formulate joint MUD and network coding as
follows: As a relay, user $i$ selects a set of users
$\mathfrak{M}_i$, and then transmits $b_m \bigoplus \dots
\bigoplus b_n$, where $m,\dots, n\in \mathfrak{M}_i$. Notice that
$\mathfrak{M}_i$ is a subset of all users that are successfully
decoded at the first stage by user $i$. At the base station, the
user's error probability is given by:
\begin{eqnarray}\label{coded_user_m}
P_r^m=P_r^{m0}\{1-(1-P_r^{mi})(1-P_r^{i0})\nonumber\\
\prod _{n\in \mathfrak{M}_i/m}[(1-P_r^{ni})(1-P_r^{n})]\},
\end{eqnarray}
and\vspace{-1mm}
\begin{equation}\label{other_user}
P_r^j\leq P_r^{j0},\ \forall j\neq \mathfrak{M}_i.\vspace{-1mm}
\end{equation}
The first term in (\ref{coded_user_m}) represents the direct
transmission error probability. The term in  brackets in
(\ref{coded_user_m}) represents the error probability from the
relay using networking coding. The successful transmission from
the relay happens only if all users in $\mathfrak{M}_i$ are
decoded correctly by user $i$, the transmission from user $i$ to
the base station is correct, and all other users are correctly
decoded at the base station. Notice that compared with
(\ref{protocol_one_BER}), the error probability for a specific
user might be worse. However, since in (\ref{coded_user_m}),
multiple users' BERs can be improved, the overall BER of the
system can be further improved under careful optimization. The
inequality in (\ref{other_user}) holds since the cancellation of
some successfully decoded users' information can improve the other
users' decoding.

The problems to be considered here are: which relays to select among the potential users;
and whose data to retransmit. We need to select the relay $i$ from
set $\mathfrak{R}$ of size $N$, and the set $\mathfrak{M}_i$ which
represents whose information should be relayed by user $i$. The problem
formulation to minimize the overall BER can be written
as:\vspace{-1mm}
\begin{equation}\label{prob_def}
\min _{\mathfrak{R},\mathfrak{M}_i} \sum_{j\in \{ K\backslash
\mathfrak{R}\}} P_r^j.\vspace{-1mm}
\end{equation}

To optimize (\ref{prob_def}), we propose an algorithm shown in
Table \ref{solution}. The basic idea is that the base station can
know at stage 1 which users' links need to be improved so as to
maximize the network performance. So at stage 2, the relay with
the best location will send the corresponding information to the
base station. At the base station, the information sent at the
first stage is delayed and combined with the relay's information
at the second stage. Consequently, the performance of all users
can be improved.

\begin{table}
\caption{Cooperative Transmission Protocols} \label{solution}
\begin{center}
\begin{tabular}{|l|}
  \hline
  1. At stage one, the relays decode using MUD.\\
  \hline
  2. At stage one, the base station decide which users are most important.\\
\hline
  3. Using feedback from the base station, at stage two, the relays \\
 \ \ \ forward the selected users' information or encode using  \\
\ \ \  network coding so as to optimize the decoding goal. \\
\hline
  4. At stage 2, MUD decoding is used at the base station. \\
  \hline
\end{tabular}
\end{center}\vspace{-5mm}
\end{table}

\vspace{-3mm}

\section{Performance Analysis\label{sec:analysis}}

In this section, we first examine the diversity order and coding
gain of the proposed protocols. Then, a performance upper bound is
given using MIMO-MUD. Finally, we study a special case for how
the relay changes the MUD asymptotic efficiency.

At stage one, we order the received signals at the base station
according to the signal strength, where user $K$ has the highest SINR
(i.e. the lowest BER). For Protocol One, we assume all relays
select user $K$'s information to retransmit if the relay decodes it
correctly. The diversity order for user $K$ can be easily shown to
be $N+1$. Since only user $K$'s copy of the information at stage 1
is retransmitted, the diversity order of the other user is still
$1$. However, the remaining users have better performance since the
strongest interference (user $K$'s signal) can be more
successfully cancelled. Define the coding gain $\rho_i$ as the
SINR improvement ratio for the remaining users. For successive
cancellation MUD, we have
\begin{equation}
\rho_{K-1}=\frac{\sigma^2+ \frac 1 M\sum_{j=1}^{K-2} A_j^2+ \frac
4 M A_{K}^2 P_r^K}{\sigma^2+ \frac 1 M\sum_{j=1}^{K-2} A_j^2+
\frac 4 M A_{K}^2 \hat P_r^K},
\end{equation}
where $\hat P_r^K$ is user $K$'s new BER and $\hat P_r^K\approx
(P_r^K)^{N+1}$. If $\frac 1 M A_K^2 >> \sigma^2+ \frac 1 M
\sum_{j=1}^{K-2} A_j^2$, the coding gain can be quite
large. For the coding gains of other users, we can calculate
$P_r^{K-1},\dots, P_r^1$ recursively. For optimal MUD, the
performance is lower bounded by that of successive cancellation
MUD.

For Protocol Two, if at the second stage, the relays retransmit
the following information
\begin{equation}
z_i=\bigoplus b_j, j\in \mathfrak{M}_i.
\end{equation}
We consider any user $i$'s information. When the SINRs are
sufficiently large, the channels between the senders and relays
are approaching ideal links. If the other direct links are also
sufficiently good, at the second stage after network decoding, the
only signals remaining are user $i$'s information from the $N$
relays, and these signals will be combined with the direct
transmission sent at the first stage. So the diversity order is
$N+1$.

Next, the proposed cooperative transmission protocol with MUD has
a performance upper bound given  by that of MIMO MUD
\cite{HuaiyuDai} in which there is infinite bandwidths between the
relays and the base station. Here we assume that the combination
is performed after decoding. Decoding error happens when all the
$N+1$ links fail, i.e.\vspace{-2mm}
\begin{equation}
P_r^k= P_r^{k0} \prod _{i=1}^N P_r^{ki},\vspace{-1mm}
\end{equation}
where $P_r^{k0}$ is the BER for direct transmission and
$P_r^{ki}$ is the BER for transmission from user $k$ to relay $i$. For
MIMO MUD, the diversity order is $N+1$.

Finally, we study a special case of two users and one relay to
investigate the performance improvement of MUD.
Here we make the approximations that the relay can always decode
correctly and the base station can use maximal ratio combining of
the direct and relay transmissions. In this ideal case, the
multiuser efficiency of optimal MUD can been expressed
as\vspace{-1mm}
\begin{eqnarray}
\eta _1\approx \min \left\{ 1, 1+\frac {(A_2+A_r)^2}{A_1^2}-2|\rho
| \frac {A_2+A_r}{A_1},\right.\nonumber \\ \left. 1+\frac
{A_2^2}{(A_1+A_r)^2}-2|\rho | \frac
{A_2}{A_1+A_r}\right\}, \vspace{-1mm}
\end{eqnarray}
where $\rho$ is the cross-correlation between the two users' waveforms,
 and $A_1$, $A_2$, and $A_r$
are the channel gains to the base station for user $1$, user $2$
and the relay, respectively.
%
%
%

\begin{figure}[htbp]
\begin{center}
    \epsfig{file=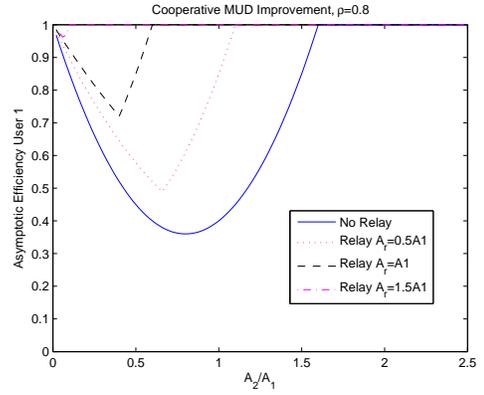,width=71truemm}
\end{center}
\caption{MUD Asymptotic Efficiency
Improvement}\label{MUD_efficiency}\vspace{-5mm}
\end{figure}

In Figure \ref{MUD_efficiency}, we show the MUD efficiency with
$A_1$=1 and $\rho=0.8$. With cooperative transmission, the
difference between the two users' SINRs can be increased so that the
multiuser efficiency can be increased. We can see that when the
relay is close to the base station (i.e. $A_r$ is large), the
multiuser efficiency can be almost $1$. We notice that this
comparison is unfair, since the bandwidth is increased with the
presence of the relay. However, when the number of users is
sufficiently larger than the number of the relays, this increase
is negligible.

\vspace{-1mm}


\section{Simulation Results}\label{sec:simulation}

The following setting is used in the simulation. We consider a
one-dimensional model where a base station, a relay, and users are
located along a line. The base station is located at 0, the two
users are located at 4 and 6, and the relay can move from 0.5 to
3.5. The  power received from a given transmitter is proportional to
$P_t/d^3$, where $d$ is the distance between the transmitter and
the receiver, and $P_t$ is the transmitted power. In the
simulation, we assume that all users and the relay use the same
transmitted power, i.e., there is no power control. We also assume
the receivers have the same additive noise level.

In the simulations, we choose a simplified model for the relay.
The relay can receive and transmit at the same time. This can be
achieved by using time sharing, different frequency bands, or
simply two relay users. We choose this model to simplify the
scenario. In each time slot, the relay
transmits a bit generated according to the protocols developed in
Section \ref{sec:Protocol}. The relay and the users transmit their
signals using CDMA, and they all use different spreading codes.
The relay and the base station perform successive-cancellation
multiuser detection on the
received signals.

\begin{figure}[htbp]
\begin{center}
    \epsfig{file=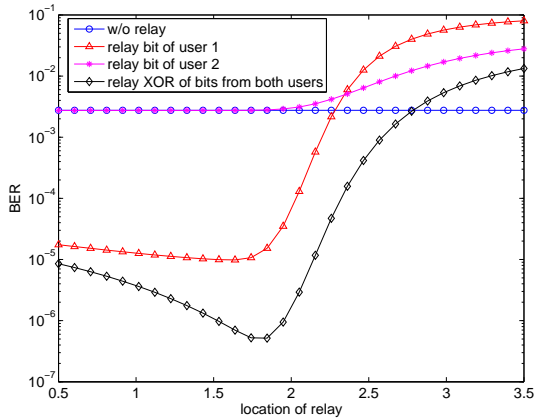,width=71truemm}
\end{center}\vspace{-2mm}
\caption{The average BER as a function of the location of the
relay. This is a high average SINR case.}
\label{fig:mean_Pe_vs_relaylocation_high_SNR}\vspace{-3mm}
\end{figure}

\begin{figure}[htbp]
\begin{center}
    \epsfig{file=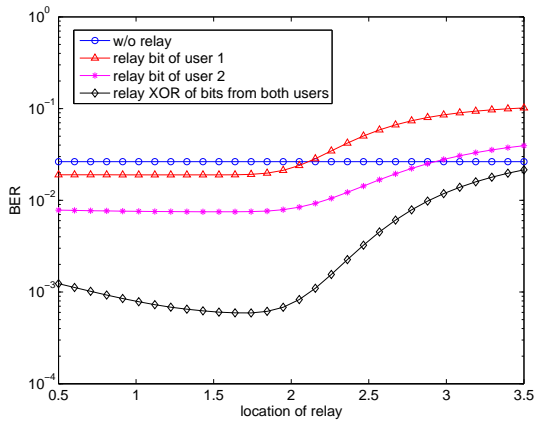,width=71truemm}
\end{center}\vspace{-2mm}
\caption{The average BER as a function of the location of the
relay. This is a low average SINR case.}
\label{fig:mean_Pe_vs_relaylocation_low_SNR}\vspace{-1mm}
\end{figure}

Figure \ref{fig:mean_Pe_vs_relaylocation_high_SNR} shows the
average BER as a function of the relay location. There are two
users, and both users and the relay use high transmitted power.
The curves correspond to the cases without the relay, with the
relay re-transmitting user 1's (located at 4) symbol, with the relay
re-transmitting user 2's (located at 6) symbol, and with the relay
re-transmitting the XOR of both users' symbols (network coding).
The first observation is that the location of the relay plays a
vital role in the system performance. The relay helps the system
performance only when its distance from the base station is below
2.5. The relay will harm the performance if it is close to the
user group. This is because for successive cancellation MUD, the
performance is better if the received power of the users is
different from each other. A relay that is close to the user group
is acting more as an interference source than as a relay. The
second observation is that there is a ``sweet spot'' for the
location of the relay around 1.6. This is because the relay's
decoding performance drops if it is located too far away from the
sources. The third observation is that the network coding protocol
with the relay re-transmitting the XOR of both users' symbols always
performs better than that when the relay just re-transmits one
user's symbol. Figure \ref{fig:mean_Pe_vs_relaylocation_low_SNR} is
similar to Figure \ref{fig:mean_Pe_vs_relaylocation_high_SNR}
except that the transmitted power is low here. We observe performance
behavior similar to the high transmitted power case.

\begin{figure}[htbp]
\begin{center}
    \epsfig{file=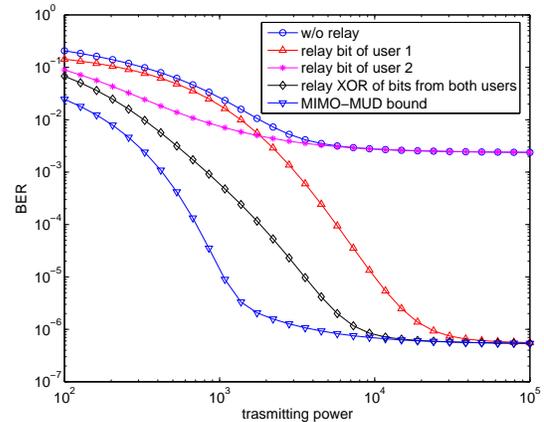,width=71truemm}
\end{center}\vspace{-1mm}
\caption{The average BER as a function of the transmitted power.}
\label{fig:mean_Pe_vs_Tpower_relay1p6}\vspace{-3mm}
\end{figure}

Figure \ref{fig:mean_Pe_vs_Tpower_relay1p6} shows the average BER
as a function of the transmitted power of the users and the relay.
The relay's location is fixed at 1.6. We can clearly see the
higher diversity order of BER vs. power for the proposed
protocols. When the transmitted power is sufficiently high, the
limiting factor of the performance is the interference. And that
is why the curve flattens when the transmitted
power grows. We notice the large difference in performance between
the case with the relay and the case without. Another interesting
observation is that, in a certain transmitted power range, relaying the
first user's symbol is better, while in other transmitted power
range, relaying the second user's symbol is better. Relaying the XOR
of both users' symbols is always the best protocol, but this requires
the use of MUD at the relays. We also show the performance of
MIMO-MUD, which serves as a performance bound.

\begin{figure}[htbp]
\begin{center}
    \epsfig{file=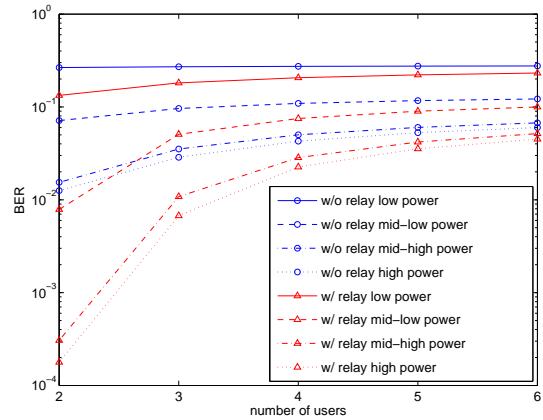,width=82truemm}
\end{center}\vspace{-1mm}
\caption{The average BER as a function of the number of
users.}\vspace{-0mm} \label{fig:mean_Pe_vs_num_user}
\end{figure}

Figure \ref{fig:mean_Pe_vs_num_user} shows the average BER as a
function of the number of users. Here we explore the cases with
two to six users. In each case, the users are uniformly
distributed in the range $[4,8]$. The relay is located at $1.6$,
and it transmits the XOR of the nearest two users' symbols. As
expected, the performance is best when there are only two users.
The performance for the case with more users can be improved by
introducing more relays or having the relay transmitting the XOR of
more users' symbols.

\vspace{-1mm}
\section{Conclusions}\label{sec:conclusion}\vspace{-1mm}

We have proposed two new cooperative transmission protocols
considering MUD as well as network coding. The enhancement of some
users' transmissions by cooperative transmission can improve the
other users' performance in MUD. Moreover, network coding can
provide additional  coding gain. From the analysis and simulation
results, the proposed protocols achieve much lower average BER,
higher diversity order and coding gain, and better asymptotic
efficiency, compared to cooperative transmission for single user
detection and traditional MUD.\vspace{-2mm}

\bibliographystyle{IEEE}

\end{document}